\documentclass[aps,prb,twocolumn,superscriptaddress,showpacs]{revtex4} 

\usepackage[dvips]{graphicx}
\usepackage{bm}

\usepackage{color}
\usepackage{latexsym} 
\usepackage{amsmath}
\usepackage{amssymb}

\begin{document}
 
\title{
Magnon instability driven by heat current in magnetic bilayers
}
 
\author{Yuichi Ohnuma}
\affiliation{Institute for Materials Research, Tohoku University, Sendai 980-8577, Japan}
\affiliation{Advanced Science Research Center, Japan Atomic Energy Agency, Tokai 319-1195, Japan} 

\author{Hiroto Adachi}
\affiliation{Advanced Science Research Center, Japan Atomic Energy Agency, Tokai 319-1195, Japan}
\affiliation{ERATO, Spin Quantum Rectification Project, Japan Science and Technology Agency, Sendai 980-8577, Japan}

\author{Eiji Saitoh}
\affiliation{Institute for Materials Research, Tohoku University, Sendai 980-8577, Japan}
\affiliation{Advanced Science Research Center, Japan Atomic Energy Agency, Tokai 319-1195, Japan}
\affiliation{ERATO, Spin Quantum Rectification Project, Japan Science and Technology Agency, Sendai 980-8577, Japan}
\affiliation{WPI Research Center, Advanced Institute for Materials Research, Tohoku University, Sendai 980-8577, Japan}

\author{Sadamichi Maekawa}
\affiliation{Advanced Science Research Center, Japan Atomic Energy Agency, Tokai 319-1195, Japan}
\affiliation{ERATO, Spin Quantum Rectification Project, Japan Science and Technology Agency, Sendai 980-8577, Japan}

\date{\today}

\begin{abstract} 
We theoretically demonstrate that, in a ferromagnet/paramagnet bilayer, a magnon instability accompanied by a gigahertz microwave emission can be driven simply by means of a temperature bias. Employing many-body theory for investigating the effects of a phonon heat current on the magnon lifetime, we show that the magnon instability occurs upon the suppression of the Umklapp scattering at low temperatures, leading to microwave emission. The present finding provides crucial information about the interplay of spin current and heat current. 
\end{abstract}
\maketitle 

\section{Introduction}
The concept of the spin-transfer torque has enabled the establishment of a quantum-mechanical way of controlling spin-based electronic devices~\cite{Slonczewski96,Berger96,Brataas12}. Although these spintronic devices are driven by spin-polarized electric currents, a recent trend among other important issues in spintronics is the thermal manipulation of these devices. Indeed, the spin Seebeck effect~\cite{Uchida08,Jaworski10,Uchida10,Xiao10,Adachi13} has enabled a thermal spin injection from insulating magnets into paramagnets, and the discovery has opened up a new research field known as spin caloritronics~\cite{Bauer12}. The field has continuously expanded and given rise to a number of intriguing phenomena emerging from the interplay between spin and heat~\cite{Costache11,Dejene13,An13,Sylvain13}. 

The basic building block of the spin Seebeck device is a bilayer composed of a ferromagnet and a paramagnetic heavy metal. This simple ferromagnet/paramagnet bilayer differs from the conventional spin-transfer torque device in that unlike the latter system, the former does not possess the so-called spin valve structure and thus the device structure can be much simpler. Given the increasing attention focused on the thermal effects in spintronics and the versatility of the simple ferromagnet/paramagnet bilayer, theoretical clarification of the thermal manipulation of ferromagnet/paramagnet bilayers~\cite{Lu12,Jungfleisch13,Kajiwara13,Cunha13} is of crucial importance.  

In this work, we theoretically address the issue of thermally driven microwave emission from the ferromagnet/paramagnet bilayer system. Using a diagrammatic perturbation calculation with respect to the external temperature bias as well as the interfacial magnetic interaction, we show that phonon heat current has remarkable influences on the magnon lifetime in the bilayer via the phonon drag mechanism, enabling thermal manipulation of the Gilbert damping. The theory explains the experiment reported in Ref.~\onlinecite{Lu12}, where thermal control of the Gilbert damping in a prototypical Y$_3$Fe$_5$O$_{12}$/Pt bilayer was demonstrated. Moreover, we predict that the Gilbert damping constant in the same bilayer system can become negative provided that Umklapp scattering of phonons is sufficiently suppressed at low temperatures, thereby causing an instability in the magnon system, leading to the microwave emission. 

The present paper is organized as follows. In Sec.~\ref{Sec:model}, we define our theoretical model to deal with a ferromagnet/paramagnet bilayer system and outline how the enhanced Gilbert damping in the system is calculated. In Sec.~\ref{Sec:Temp}, we introduce a temperature bias to our model and discuss its effects on the enhanced Gilbert damping. Based on this result, in Sec.~\ref{Sec:Emission}, we predict that thermally driven microwave emission from the bilayer is possible at low temperatures. Finally, in Sec.~\ref{Sec:Conclusion}, we discuss the physics behind the predicted microwave emission and then summarize our result.

\section{Model and enhanced Gilbert damping \label{Sec:model}} 
\begin{figure}[t] 
  \begin{center}
        \includegraphics[width=8.5cm]{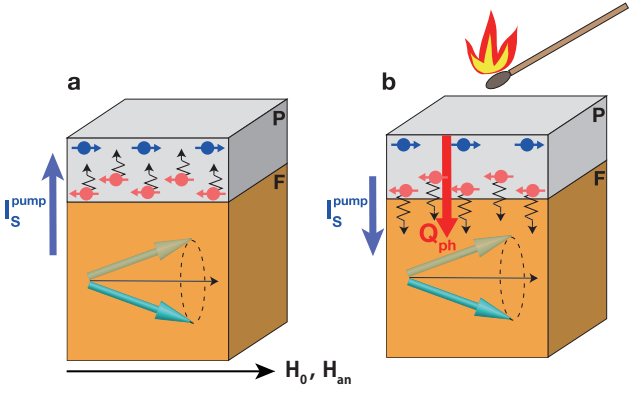} 
  \end{center}
\caption{ (color online) 
  Schematic of the magnetic bilayer and a physical picture of thermally driven microwave emission, where $H_0$ is an applied static magnetic field and $H_{\rm an}$ is the anisotropy field ($H_{\rm an} \ll H_0$). (a) In the absence of a temperature bias, a precessing magnetization in the ferromagnet $F$, or a magnon, creates spin current $I_{\rm S}^{\rm pump}$ pumped into the adjacent paramagnet, $P$. This effect always results in an enhancement of the Gilbert damping constant because the magnons lose their spin angular momenta. (b) When the paramagnet is heated, the pumped spin current tends to diffuse back into the ferromagnet because of the drag effect of the phonon heat current $Q_{\rm ph}$, and hence produces a forced spin backflow. This spin backflow reduces the Gilbert damping constant, as the magnons gain additional spin angular momenta, thereby causing a magnon instability, leading to microwave emission.} 
\label{fig1_Ohnuma}
\end{figure}

The system under consideration is a bilayer composed of an insulating ferromagnet ($F$) and a metallic paramagnet ($P$) under a temperature bias (Fig.~\ref{fig1_Ohnuma}). An external static magnetic field $H_0$ is applied in the lateral direction, and the anisotropy field $H_{\rm an}$ is assumed to be in the same direction, with its strength being much weaker than $H_0$. The physics of $F$ and $P$ are, respectively, described by the localized spins and the spin accumulation, and the interaction between them is given by the $s$-$d$ exchange coupling at the interface. Since, in our approach, the nonequilibrium dynamics of these quantities play a key role, it is convenient to employ a diagrammatic perturbation approach~\cite{Dominicis75,Michaeli09} formulated in terms of the quantum action. We use the matrix form of the propagators represented in terms of the retarded, advanced, and Keldysh components~\cite{Rammer86}. 

Our starting point is the following action:~\cite{Dominicis75,Ohnuma14} 
\begin{eqnarray}
  A_0 &=& \int_C dt \Big\{ \sum_{\bf q} a^\dag_{\bf q}(t)
  G_{\bf q}^{-1} ({\rm i} \partial_t ) a_{\bf q}(t)   \nonumber \\
  &+&\sum_{\bf k} s^+_{{\bf k}}(t) \chi_{\bf k}^{-1} ({\rm i} \partial_t ) 
  s^-_{\bf k}(t) 
  \nonumber \\
  &+& \sum_{{\bf q},{\bf k}} \big[ J_{\rm sd}({{\bf k}-{\bf q}}) 
    s_{\bf k}^+(t) a^\dag_{\bf q} (t) 
      + \text{H.c.} \big] \Big\}, 
  \label{Eq:action01}
\end{eqnarray}
where the integration is performed along the Keldysh contour $C$. The first term in Eq.~(\ref{Eq:action01}) describes the dynamics of magnons, represented by the Holstein-Primakoff operators $a_{\bf q}$ and $a^\dag_{\bf q}$. Here the retarded part of the magnon propagator is given by $G^{R}_{\bf q} (\omega)= 1/(\omega- \omega_{\bf q} + {\rm i} \alpha^{(0)} \omega)$, with $\omega_{\bf q}$ and $\alpha^{(0)}$ being the magnon excitation energy and the intrinsic Gilbert damping constant, respectively. In the following, the uniform magnon frequency is denoted as $\omega_0$. The second term in Eq.~(\ref{Eq:action01}) describes the dynamics of the paramagnetic spin density~\cite{Hertz74} (spin accumulation), represented by $s^\pm_{\bf k}= s^x_{\bf k} \pm {\rm i} s^y_{\bf k}$, and the retarded part of the propagator is given by $\chi_{\bf k}^{R}(\omega)= \chi_{P}/(1+ \lambda^2 {\bf k}^2 - {\rm i} \omega \tau_{\rm sf})$, where $\chi_{P}$ is the uniform paramagnetic susceptibility, $\lambda$ is the spin diffusion length, and $\tau_{\rm sf}$ is the spin-flip relaxation time~\cite{Fulde68}. The third term in Eq.~(\ref{Eq:action01}) describes the $s$-$d$ exchange interaction at the $F$/$P$ interface, where $J_{\rm sd}({{\bf k}- {\bf q}})$ is the Fourier transform of $J_{\rm sd} \sum_{{\bf r}_0 \in \text{interface}} v_0 \delta({\bf r} - {\bf r}_0)$ with $J_{\rm sd}$ and $v_0$ being the strength of the $s$-$d$ interaction and the cell volume, respectively.

We begin by explaining how the damping of magnons in this bilayer system, conventionally represented by the effective Gilbert damping constant, is calculated in our approach. Guided by the experimental observation that the use of a poor spin sink such as Cu makes the temperature bias effect invisible~\cite{Lu12,Cunha13}, we assume that $P$ is a good spin sink with sizable spin-orbit scattering, such as Pt. Since the attachment of a good spin sink introduces a new source of spin dissipation, the Gilbert damping constant in the bilayer is given as a sum of the intrinsic contribution $\alpha^{(0)}$ (defined merely by $F$), and an enhanced Gilbert damping constant $\delta \alpha$ caused by the attached $P$, i.e., $\alpha = \alpha^{(0)} + \delta \alpha$. As discussed in Ref.~\onlinecite{Ohnuma14}, the enhanced Gilbert damping constant is in general calculated as $\delta \alpha = - (1/\omega_0) {\rm Im} \Sigma_0^R(\omega_0)$, where $\Sigma_{0}^{R}(\omega_0)$ is the selfenergy for the uniform-mode magnon that is defined by the following Dyson equation~\cite{Doniach-text}: $1/g^R_0(\omega)= 1/G^{R}_0(\omega)-\Sigma_{0}^{R}(\omega)$ with $g^{R}_0(\omega)$ being the renormalized propagator of the uniform-mode magnon. 

\begin{figure}[t] 
  \begin{center}
    \includegraphics[width=7cm]{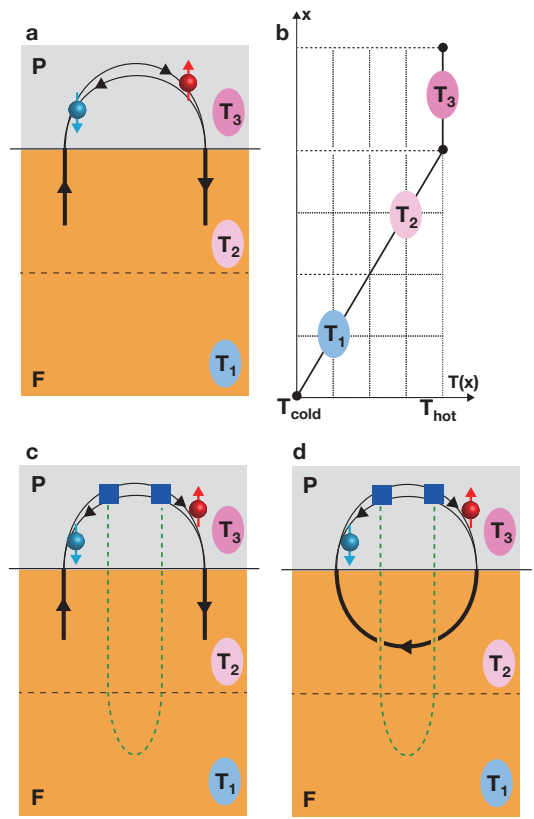}
  \end{center}
\caption{ (color online) 
 (a) Diagrammatic representation of the basic process that gives the enhanced Gilbert damping, which is free from the temperature bias effect. Here the bold solid line and the double solid lines with a pair of opposite arrows, respectively, represent the propagators of the magnon and the spin accumulation. (b) Temperature profile of the system, where the paramagnet $P$ is in contact with a heat source of temperature $T_{\rm hot}$ and the ferromagnet $F$ is in contact with a heat sink of temperature $T_{\rm cold}$. In our modeling, the continuous temperature profile (thick solid line) is replaced by three discrete local temperatures of $T_1$, $T_2$ and $T_3$. 
  (c) The phonon-drag process that gives the thermal control of the Gilbert damping. Here the dashed line represents the phonon propagator. 
  (d) The Feynman diagram for the longitudinal spin Seebeck effect discussed in Ref.~\onlinecite{Adachi13b}. }
\label{fig2_Ohnuma}
\end{figure}

The enhanced Gilbert damping constant $\delta \alpha$ is calculated from the magnon selfenergy shown in Fig.~\ref{fig2_Ohnuma}(a). Performing a perturbative approach with respect to $J_{\rm sd}$, we obtain 
\begin{eqnarray}
  \delta \alpha &=& 
  \frac{\langle \langle J^2_{\rm sd} \rangle \rangle }{\hbar^2} 
  \sum_{\bf k}  \frac{1}{\omega_0} \text{Im} \chi^R_{\bf k} (\omega_0) , 
  \label{Eq:Dalpha01} 
\end{eqnarray}
where $\langle \langle J^2_{\rm sd} \rangle \rangle =2 J^2_{\rm sd} S_0 N_{\rm int}/( N_{P} N_{F})$, with $\hbar$, $N_{\rm int}$, $N_F$, and $N_P$ being, respectively, the Planck constant divided by $2 \pi$, the number of localized spins at the interface, and the number of lattice sites in $F$ and $P$. Such enhancement of the Gilbert damping has been confirmed experimentally by linewidth measurements using ferromagnetic resonance (FMR)~\cite{Mizukami02,Heinrich03}. 

The result can be interpreted from the viewpoint of spin transfer across the interface~\cite{Silsbee79,Tserkovnyak02}, which is now termed ``spin pumping''. Because the exchange interaction at the interface conserves the total spin, the enhanced Gilbert damping, or spin angular momentum loss, should be accompanied by a spin transfer from $F$ into $P$, where additional spin dissipation occurs. Following the procedure given in Ref.~\onlinecite{Ohnuma14}, a spin current, $I_{\rm S}^{\rm pump}$, that is pumped from $F$ into $P$ under the FMR condition is shown to be intimately connected to $\delta \alpha$ through the relation 
\begin{eqnarray}
  I_{\rm S}^{\rm pump} &=& 
  \delta \alpha 
  \frac{ S_0 N_{\rm F} (\gamma h_{\rm rf})^2} 
       {(\alpha^{(0)})^2 \omega_0},
       \label{Eq:Ipump01}
\end{eqnarray}
where $\gamma$ and $h_{\rm rf}$ are, respectively, the gyromagnetic ratio and the amplitude of the oscillating microwave field. Since a magnon carries a spin angular momentum $-\hbar$, this effect pumps negative spin angular momenta into $P$ [Fig.~\ref{fig1_Ohnuma}(a)]. The existence of the pumped spin current has been demonstrated experimentally using the inverse spin Hall effect~\cite{Saitoh06}. 

The perturbation approach with respect to the $s$-$d$ interaction is justified when $J_{\rm sd}$ is of the order of several tens meV such as Y$_3$Fe$_5$O$_{12}$/Pt systems~\cite{Kajiwara10,Cornelissen15}. Indeed, we find that the expansion parameter characterizing the strength of the higher-order terms compared to the lowest-order term is given by $\epsilon=J_{\rm sd}^2 \chi_P/J_{\rm ex}$, where $J_{\rm ex}$ is the exchange integral of $F$. Because our definition of spin susceptibility differs from the usual magnetic susceptibility by a factor $\mu_{\rm B}^2$ ($\mu_{\rm B}$: the Bohr magneton), the expansion parameter is calculated to be $\chi_P \sim A_{\rm Stoner}/E_{\rm F}$, where $A_{\rm Stoner}$ and $E_{\rm F}$ are respectively the dimensionless Stoner enhancement factor and Fermi energy of $P$. Therefore we have an estimate $\epsilon \sim (J_{\rm sd}/J_{\rm ex}) (A_{\rm Stoner} J_{\rm sd}/E_{\rm F}) \ll 1$, where we have used the fact that, in the case under consideration, $J_{\rm sd}/J_{\rm ex} \sim 1$ whereas $A_{\rm Stoner} J_{\rm sd}/E_{\rm F} \ll 1$ even if we consider a moderate strength of the Stoner enhancement factor of Pt, $A_{\rm Stoner} \sim 5$~\cite{Andersen70}. This estimate justifies the perturbation approach with respect to $J_{\rm sd}$ in Y$_3$Fe$_5$O$_{12}$/Pt systems.

\section{Effects of temperature bias \label{Sec:Temp}} 
Now, we discuss the effects of a temperature bias on the enhanced Gilbert damping. We consider a situation where $F$ and $P$ are in contact with heat baths with temperatures $T_{\rm cold}$ and $T_{\rm hot}$, and where the thermal resistivity and the thickness of $P$ are considerably smaller than those of $F$, such that the temperature profile of the bilayer behaves as shown in Fig.~\ref{fig2_Ohnuma}(b). It is well known that the temperature is a statistical property of the system; therefore, no Hamiltonian exists to describe the effects of a temperature bias~\cite{Kubo-text,Luttinger64}. To consider the thermal effects in our approach, a model is introduced in which the system is divided into three temperature domains with local equilibrium temperatures of $T_1=(T_{\rm hot}+ 3T_{\rm cold} )/4$, $T_2=(3T_{\rm hot}+ T_{\rm cold})/4$, and $T_3=T_{\rm hot}$. We assume that each domain is initially in local thermal equilibrium without interaction with the neighboring domains, and then switch on the interactions between them, and calculate the nonequilibrium dynamics of the system. This method has been proven to be quite efficient for a theoretical description of the spin Seebeck effect~\cite{Adachi13}. 

The size of the temperature domains may be given by a length scale defining local equilibrium temperature, which is set by the mean free path of acoustic phonons~\cite{Mahan-text}. In the case of a widely used Y$_3$Fe$_5$O$_{12}$, if we neglect so-called subthermal effects by which low-energy long-ranged phonons carry most of the heat, it is of the order of $1$ nm at $300$ K and $1$ $\mu$m at $50$ K~\cite{Boona14}. If we take into account the subthermal effects, the length is expected to be much longer. Apart from this estimate, however, the important point of the following calculation is that the result scales with the temperature gradient, instead of the temperature difference itself [see the paragraph below Eq.~(\ref{Eq:c5})]. Thus, the following result is independent of the choice of the domain size. 

We apply this technique to the present problem and calculate the enhanced Gilbert damping constant. We then find that, as long as model (\ref{Eq:action01}) consists merely of magnons and spin accumulation, there is no temperature bias effect on the enhanced Gilbert damping, even though there is a finite temperature jump $T_3 - T_2 \neq 0$ at the $F$/$P$ interface in our model calculation. That is, even if we evaluate the diagram in Fig.~\ref{fig2_Ohnuma}(a) under a temperature bias ($T_{\rm hot} \neq T_{\rm cold}$), no effects arise from this temperature bias. This result originates from the fact that the enhanced Gilbert damping is a purely ``mechanical'' process~\cite{Luttinger64} that is not represented by magnon distribution function. For the thermal bias to play a role, we need to consider a process that involves, in addition to the magnons and spin accumulation, a third degree of freedom that carries heat and interacts with the magnons or spin accumulation. 

To consider the temperature bias effects, we thus focus on acoustic phonons as the third degree of freedom. In a widely used Y$_3$Fe$_5$O$_{12}$/Pt bilayer system, it is well established that acoustic phonons give a quite sharp peak to the spin Seebeck effect, owing to the phonon-drag mechanism~\cite{Adachi10,Jaworski11,Uchida12,Tikhonov13} at low temperature. Following Ref.~\onlinecite{Adachi10}, we consider the following action: 
\begin{eqnarray}
  A_\text{ph} &=& \int_C dt \Big\{ \sum_{\bf k} B^\dag_{\bf k}(t) 
  D_{\bf k}^{-1} ({\rm i} \partial_t) B_{{\bf k}}(t) \nonumber \\
  &&\hspace{1cm}
  + \sum_{{\bf k},{\bf K}} \Upsilon_{{\bf K}} B_{\bf K} (t) 
  s^+_{{\bf k}+{\bf K}}(t) s^-_{\bf k}(t) \Big\}, 
  \label{Eq:action_ph}
\end{eqnarray}
where $B^\dag_{\bf K}= b^\dag_{\bf K}+ b_{\bf -K}$ is the creation operator for phonons, and $D^R_{\bf K}(\nu)=(\nu- \nu_{\bf K} + {\rm i}/\tau_\text{ph})^{-1} - (\nu+ \nu_{\bf K}+ {\rm i}/\tau_\text{ph})^{-1}$ is the retarded part of the phonon propagator, with $\tau_{\rm ph}$ and $\nu_{\bf K}= v_{\rm ph} K$ being the phonon lifetime and energy; here the phonon propagator depends on the underlying material. The first term describes the dynamics of the acoustic phonon, and the second term describes the interaction between the phonon and the spin accumulation in $P$, where the explicit expression of the vertex is given by $\Upsilon_{{\bf K}}=(g_{\rm s \mathchar`-p}/\hbar) \sqrt{\hbar \nu_{\bf K}/2 M_{\rm ion} v^2_{\rm ph}}$ with $M_{\rm ion}$ and $g_{\rm s \mathchar`-p}$ being the ion mass and the strength of the spin accumulation-phonon interaction~\cite{Adachi13b}, respectively. Note that in Eq.~(\ref{Eq:action_ph}), effects of the phonon-phonon Umklapp scattering is phenomenologically taken into account through the temperature dependence of the phonon lifetime $\tau_{\rm ph}$, which is know to be largely suppressed at room temperature by Umklapp processes. Besides, the spin accumulation-phonon Umklapp scattering is ill-defined because the spin accumulation is a hydrodynamic diffusive mode. Therefore, Eq.~(\ref{Eq:action_ph}) is valid even at room temperature.

In the following, we show that, in the presence of phonon heat current concomitant to the temperature bias, the total Gilbert damping $\alpha$ can be written as 
\begin{eqnarray}
  \alpha &=& \alpha^{(0)} + \delta \alpha + \delta \alpha' \Delta T, 
  \label{Eq:alpTOT01} 
\end{eqnarray}
where the last term represents the thermally driven change in the Gilbert damping. Notably, as shown below, the term $\delta \alpha' \Delta T$ is proportional to the phonon heat current, $Q_{\rm ph}$, a situation that is accounted for only by nonequilibrium phonon dynamics. 

The dominant contribution to $\delta \alpha' \Delta T$ at low temperature comes from the process shown in Fig.~\ref{fig2_Ohnuma}(c), where the acoustic phonons carry the heat current $Q_{\rm ph}$ and drag the spin accumulation. 
As before, we have a relation $\delta \alpha' \Delta T= - (1/\omega_0) {\rm Im} \Sigma_{0}^{'R}(\omega_0)$ for the present problem, where 
\begin{eqnarray}
  \Sigma_{0}^{'R}(\omega) &=& 
  \frac{J_{\rm sd}^2S_0 N_{\rm int}^{(2,3)}}{\sqrt{2}N_F N_P^2} \sum_{{\bf k}, {\bf K}_3} 
  \Upsilon^2_{{\bf K}_3}[\chi_{\bf k}^R(\omega)]^2 \nonumber \\
  &\times& {\rm i} \int_\nu \chi_{{\bf k}- {\bf K}_3} (\omega- \nu)
  \delta D_{{\bf K}_3}^K(\nu) 
\end{eqnarray}
is the magnon selfenergy corresponding to the diagram in Fig.~\ref{fig2_Ohnuma}(c), and $ \delta D_{{\bf K}_3}^K(\nu)$ is given by 
\begin{eqnarray}
  \delta D_{{\bf K}_3}^K(\nu) &=& 
  \frac{L}{N_F^2} \sum_{{\bf K}_1,{\bf K}_2} |D^R_{{\bf K}_3}(\nu)|^2 |D^R_{{\bf K}_2}(\nu)|^2 
       {\rm Im} D^R_{{\bf K}_1}(\nu) \nonumber \\
       && \times 
       \left[ \coth \big( \frac{\hbar \nu}{2 k_{\rm B}T_3} \big) 
         -\coth \big( \frac{\hbar \nu}{2 k_{\rm B}T_1} \big)\right]. 
       \label{Eq:dD_K01}
\end{eqnarray}
We evaluate the integral over $\nu$ by picking up the dominant condribution proportional to $\tau_{\rm ph}$, and obtain 
\begin{eqnarray}
  \Sigma_{0}^{'R}(\omega) &=& 
  \frac{J_{\rm sd}^2S_0 N_{\rm int}^{(2,3)}}{N_F^3 N_P^2} L \tau_{\rm ph} 
  {\rm i} \sum_{{\bf k},{\bf K}_1,{\bf K}_2,{\bf K}_3} \Upsilon^2_{{\bf K}_3}
       [\chi_{\bf k}^R(\omega)]^2  \nonumber \\
       &\times& 
       [\chi^R_{{\bf k}_-}(\omega_-)+ \chi^R_{{\bf k}_+}(\omega_+)]
       {\rm Im} D_{{\bf K}_3}^R(\nu_{{\bf K}_2})] \nonumber \\
       &\times& 
         \left[ \coth \big( \frac{\hbar \nu_{{\bf K}_2}}{2 k_{\rm B}T_3} \big) 
         -\coth \big( \frac{\hbar \nu_{{\bf K}_2}}{2 k_{\rm B}T_1} \big)\right], 
         \label{Eq:dSigma01}
\end{eqnarray}
where ${\bf k}_\pm= {\bf k} \pm {\bf K}_1$, $\omega_\pm = \omega \pm \nu_{{\bf K}_2}$, and we have used the relation ${\rm Im}D_{\bf K}^R(-\nu)= -{\rm Im}D_{\bf K}^R(\nu)$. Then, after a tedious but straightfoward calculation, we have 
\begin{eqnarray}
  \delta \alpha' \Delta T
  &=& 
  \frac{6 \pi}{\hbar} \delta \alpha \chi_P^2 \widetilde{\Gamma}_{\rm eff}^2 L \tau_{\rm ph} 
  \frac{\big( \rho^F_{\rm ph}(\nu_{F}) \big)^2 (k_{\rm B}T)^5 }
       {(\hbar \nu_{F})^4}  %
       \Big( \frac{\Delta T}{T} \Big) c_5, \nonumber \\
       \label{Eq:Dalpha_micro}
\end{eqnarray}
where $\nu_F$ is the Debye frequency of $F$,  
\begin{eqnarray}
  c_5 &=& \int_0^{\hbar \nu_F/k_{\rm B}T} dx \frac{x^5}{4 \sinh^2(x/2)}, 
  \label{Eq:c5}
\end{eqnarray}
and the quantity $L$ is defined below Eq.~(\ref{Eq:Qph00}) in Appendix~\ref{Sec:Append1}. 

Note that the above result [Eq.~(\ref{Eq:Dalpha_micro})] scales with the temperature gradient instead of the temperature difference itself, which can be seen from the fact that $\delta \alpha$ and $L$ in Eq.~(\ref{Eq:Dalpha_micro}) are, respectively, proportional to $1/N_F$ and $N_{\rm int}^{(2,3)}$. Because $N_F = L_x L_y L_z/ a^3$ and $N_{\rm int}^{(2,3)}= L_y L_z / a^2$ where $a$ is the lattice spacing and $L_x L_y L_z$ is the volume of $F$ with the $x$-axis chosen along the temperature gradient, Eq.~(\ref{Eq:Dalpha_micro}) is proportional to $a (\Delta T/L_x) = a \nabla T$. 

The right hand side of Eq.~(\ref{Eq:Dalpha_micro}) can be rewritten in terms of phonon heat current $Q_{\rm ph}$ across the $F/P$ interface. Using Eqs.~(\ref{Eq:Qph01}) and (\ref{Eq:Qph02}) in Appendix~\ref{Sec:Append1}, the thermal contribution to the enhanced Gilbert damping can be expressed as 
\begin{eqnarray}
  \delta \alpha' \Delta T&=& 
  \frac{12}{\pi^2} \chi_{P}^2 \widetilde{\Gamma}^2_{\rm eff} B \delta \alpha \, \, Q_{\rm ph}, 
  \label{Eq:Dalpha02}
\end{eqnarray}
where 
$\widetilde{\Gamma}^2_{\rm eff}= (g_{\rm s \mathchar`- p}^2 /\hbar M_{\rm ion}v_{\rm ph}^2)N^{-1}_{\rm N} \sum_{\bf K} \nu_{\bf K}/(\nu_{\bf K}^2+ 1/\tau^2_{\rm ph})$ 
is the effective coupling constant between the phonon and the spin accumulation, and a dimensionless coefficient $B$ is given by 
\begin{eqnarray}
  B
  &=&
  \frac{\big( \hbar \nu_{P} \big)^6 \big( \rho^F_{\rm ph}(\nu_F) \big)^2}
       {\big( \hbar \nu_{F} \big)^4 \rho^P_{\rm ph}(\nu_P) \big( \rho^F_{\rm ph}(\nu_P) \big)^2} 
       \frac{c_5/c_8}{(k_{\rm B}T)^3}. 
\end{eqnarray}

Equation (\ref{Eq:Dalpha02}) is the main result of the present paper. As introduced before, the quantity $Q_{\rm ph}= -K_{\rm ph} \Delta T$ in Eq.~(\ref{Eq:Dalpha02}) represents the phonon heat current under the temperature bias $\Delta T= T_{\rm hot}- T_{\rm cold}$, where the expression for $K_{\rm ph}$ ($>0$) is given by Eq.~(\ref{Eq:Qph02}). Note that, within a linear response with respect to $\Delta T$, the phonon-drag process discussed here gives the leading contribution to $\delta \alpha'$ at low temperature, which also well explains the phonon-drag peak in the thermal conductivity and the spin Seebeck effect in LaY$_2$Fe$_5$O$_{12}$~\cite{Adachi10}. 

\begin{figure}[t] 
  \begin{center}
    \includegraphics[width=7cm]{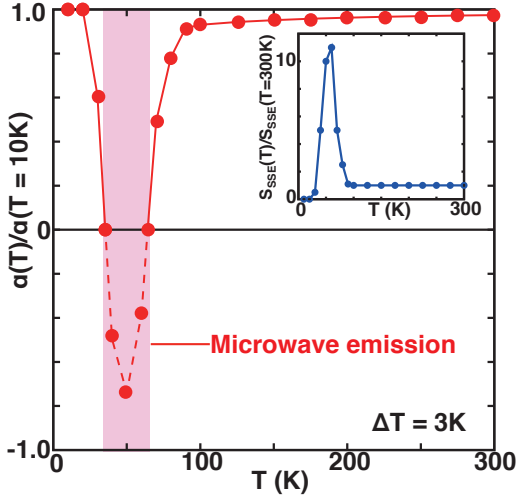} 
  \end{center}
\caption{ (color online) 
  The temperature dependence of the total Gilbert damping constant $\alpha= \alpha^{(0)}+ \delta \alpha+ \delta \alpha' \Delta T$ calculated for a Y$_3$Fe$_5$O$_{12}$/Pt bilayer under a temperature bias of $\Delta T= 3$ K. Here the non-thermal terms $\alpha^{(0)}$ and $\delta \alpha$ are taken from Ref.~\onlinecite{Lu12}, while the thermally driven term $\delta \alpha' \Delta T$ is calculated using Eq.~(\ref{Eq:Dalpha03}). Inset: The temperature dependence of the spin Seebeck signal $S_{\rm SSE}$ taken from Ref.~\onlinecite{Uchida12}, which is used to calculate $\delta \alpha' \Delta T$ via Eq.~(\ref{Eq:Dalpha03}). 
}
\label{fig3_Ohnuma} 
\end{figure}

\section{Microwave emission driven by temperature bias \label{Sec:Emission}} 
The most remarkable prediction of the present theory is that the total Gilbert damping constant $\alpha$ in Eq.~(\ref{Eq:alpTOT01}) can become negative, signifying an instability in the magnon system and giving rise to gigahertz microwave emission. The prediction is made as follows. First, according to Eq.~(\ref{Eq:Dalpha02}), the sign of $\delta \alpha' \Delta T$ can be negative when $P$ is hotter because of the condition that $Q_{\rm ph}=-K_{\rm ph}(T_{\rm hot} - T_{\rm cold}) <0$. Second, the magnitude of $\delta \alpha' \Delta T$ is significantly enhanced at low temperatures where the phonon thermal conductivity $K_{\rm ph}$ has a phonon-drag peak~\cite{Slack71}, thus compensating for the positive contribution from $\alpha^{(0)}+ \delta \alpha$ and leading to a negative Gilbert damping constant. 

To judge the feasibility of the predicted microwave emission, we calculate the temperature dependence of $\alpha$ using information given in the literature. In doing so, we first limit ourselves to a typical Y$_3$Fe$_5$O$_{12}$/Pt bilayer system~\cite{Lu12,Jungfleisch13,Kajiwara13,Cunha13}, in which the reduction of the Gilbert damping constant under a temperature bias is actually observed. We next use the fact that the longitudinal spin Seebeck effect in the same system is governed by a quite similar phonon-drag process~\cite{Uchida12,Adachi13b}, shown in Fig.~\ref{fig2_Ohnuma}(d). 
Use of the longitudinal spin Seebeck signal allows us to eliminate experimentally unknown parameter $\chi_P^2 \widetilde{\Gamma}^2_{\rm eff}$ from Eq.~(\ref{Eq:Dalpha02}). 

We can calculate the spin current generated by the longitudinal spin Seebeck effect, $I_{\rm S}^{\rm SSE}$, following the procedure given in Ref.~\onlinecite{Adachi13b} (see Appendix~\ref{Sec:Append2} for details). The result is 
\begin{eqnarray}
  I_{\rm S}^{\rm SSE} &=& 
  \delta \alpha N_F 4 \pi^2 L \tau_{\rm ph} 
  \widetilde{\Gamma}^2_{\rm eff} 
  \frac{\chi_P^2 \tau_{\rm sf}^2}{\hbar^5} 
  \frac{\rho_{\rm mag}(\omega_{\rm D})}{\sqrt{\hbar \omega_{\rm D}}} \nonumber \\
  && \times 
  \frac{\big(\rho^F_{\rm ph}(\nu_F)\big)^2 (k_{\rm B}T)^{19/2}}{(\hbar \nu_F)^4}
  \left( \frac{\Delta T}{T} \right) c_{3/2}\, c_6, 
  \label{Eq:ISSE_micro}
\end{eqnarray}
where $\rho_{\rm mag}(\omega)$ is the density of states of magnon, $\omega_{\rm D}= 2 \pi^2 J_{\rm ex}S_0/\hbar$ with $J_{\rm ex}$ being the exchange energy of $F$, and $c_{3/2}$ and $c_6$ are respectively defined by 
\begin{eqnarray}
  c_{3/2} &=& 
  \int_{0}^{\hbar\omega_D/k_{\rm B}T} dx \frac{x^{3/2}}{\sinh(x/2)}, \\
  c_{6} &=& 
  \int_{0}^{\hbar \nu_F/k_{\rm B}T} dx \frac{x^6}{4 \sinh(x/2)}. 
\end{eqnarray}
By comparing Eq.~(\ref{Eq:Dalpha_micro}) with Eq.~(\ref{Eq:ISSE_micro}), we finally obtain 
\begin{eqnarray}
\delta \alpha'  (T) \Delta T&=& (1/\omega_0) C(T) \, I_{\rm S}^{\rm SSE}(T), 
\label{Eq:Dalpha03}
\end{eqnarray}
where a dimensionless coefficient $C(T)$ is given by 
\begin{eqnarray}
  C(T) &=& 
  \frac{3}{\pi} 
  \frac{ \omega_0/\omega_{\rm D}^2  }
       {\rho_{\rm mag}(\omega_{\rm D})(\omega_{\rm D} \tau_{\rm sf})^2} 
  \Big(\frac{\hbar\omega_{\rm D}}{k_{\rm B} T}\Big)^{9/2}  
  \frac{c_{5}}{c_{3/2} \, c_{6}}. 
\end{eqnarray}
Note that the above result is derived within the linear-response calculation and thus the result is always linear with respect to the temperature bias $\Delta T$. 

An important consequence of Eq.~(\ref{Eq:Dalpha03}) is that it gives the relation 
\begin{equation}
\delta \alpha' (T) \Delta T
= \delta \alpha'(T_{\rm R})\Delta T 
\left( \frac{C(T)}{C(T_{\rm R})} \right) 
\left( \frac{I_{\rm S}^{\rm SSE}(T)}{I_{\rm S}^{\rm SSE}(T_{\rm R})} \right), 
\end{equation}
where $T_{\rm R}$ is the room temperature. From this, we can estimate $\delta \alpha' (T) \Delta T$, because the temperature dependence of the ratio $I_{\rm S}^{\rm SSE}(T)/I_{\rm S}^{\rm SSE}(T_{\rm R})$ is given in Ref.~\onlinecite{Uchida12}, the value of $\delta \alpha'(T_{\rm R}) \Delta T$ is known from Ref.~\onlinecite{Lu12}, and the value of $C(T)/C(T_{\rm R})$ can be calculated numerically. The validity of this procedure is firstly checked by comparing it with an experimental result reported in Ref.~\onlinecite{Lu12}, where a 17\% reduction in the Gilbert damping constant in a Y$_3$Fe$_5$O$_{12}$/Pt bilayer is observed at room temperature when the Pt side is hotter by an amount $\Delta T=20$K (see Appendix~\ref{Sec:Append3}). 

Figure \ref{fig3_Ohnuma} shows the temperature dependence of the total Gilbert damping constant $\alpha$, calculated for a Y$_3$Fe$_5$O$_{12}$/Pt hybrid system, where the non-thermal terms $\alpha^{(0)}$ and $\delta \alpha$ are taken from Ref.~\onlinecite{Lu12} and assumed to be temperature-independent. Clearly, we find that a negative Gilbert damping constant is achieved at around $T=50$ K under a temperature bias of $\Delta T=3$ K, which signifies an instability in the uniform-mode magnon dynamics. 

This magnon instability with the negative total Gilbert damping constant indicates that the energy gain due to the work done by the temperature bias exceeds the energy loss due to the non-thermal Gilbert damping constant, $\alpha^{(0)}+ \delta \alpha$, such that the excess energy dissipates in the form of emitted microwave radiation. This is similar to the case of microwave emission from spin valves that are driven by a spin-transfer torque~\cite{Kiselev03,Rippard04}. Unlike spin valve systems, however, magnetization reversal does not occur; for the magnetization reversal to be realized, two well-defined energy minima in the free energy---corresponding to parallel and anti-parallel states in the case of the spin valve---are required. In the present case, on the other hand, because we are dealing with a situation where the external magnetic field is considerably larger than the in-plane anisotropy field as stated in the beginning of this section, the system possesses only one energy minimum, which corresponds to the direction of the external magnetic field. In such a high-field region, the magnon instability signifies only microwave emission, even in the spin valve system~\cite{Kiselev03}. 

Therefore, this calculation shows that we can drive a gigahertz microwave emission with work done by an applied temperature bias. Because the uniform mode has the narrowest linewidth in the spin-wave resonance spectrum in the Y$_3$Fe$_5$O$_{12}$/Pt system~\cite{Sandweg10}, the instability is expected to originate from uniform precession. However, since a magnetostatic surface mode may have stronger coupling to the Pt layer because of its surface-localized nature, an instability of this mode prior to the uniform precession may also be possible. Although the microwave emission in an actual experiment is expected to be concomitant with the dynamically precessing state involving highly nonlinear processes, the present theory deals only with linear stability analysis; the precise description of such a dynamically precessing state is beyond our scope. 

It is important to note that the predicted microwave emission can be separated from background thermal radiations by measuring the power spectra of emitted microwaves~\cite{Kajiwara10,Collet15}. The predicted microwave emission has peaks only at several gigahertz frequencies, which is specific to magnon dynamics (see Fig. 3 in Ref.~\onlinecite{Kajiwara10}). By contrast, the background thermal radiations do not possess such features. 

Before ending this section, it is informative to comment on a key point to experimentally achieve the predicted microwave emission. In experiments, an Y$_3$Fe$_5$O$_{12}$/Pt bilayer is conventionally grown on a millimeter-thick Gd$_3$Ga$_5$O$_{12}$ substrate. The thickness of the Y$_3$Fe$_5$O$_{12}$ film is less than a micrometer in order for the effect of the spin transfer across the Y$_3$Fe$_5$O$_{12}$/Pt interface to be visible. On the other hand, in our calculation we assume that, at low temperatures, the phonon mean free path grows to a millimeter scale (recall that in Ref.~\onlinecite{Uchida12}, from which we take the input data to draw Fig.~\ref{fig3_Ohnuma}, a bulk Y$_3$Fe$_5$O$_{12}$ was used). Therefore, an implicit assumption in our theory is that the interface between the Y$_3$Fe$_5$O$_{12}$ film and the Gd$_3$Ga$_5$O$_{12}$ substrate is sufficiently transparent to phonons. This means that, although the quality of Y$_3$Fe$_5$O$_{12}$/Pt interface is usually the only matter of attention, much caution should be paid to the quality of Y$_3$Fe$_5$O$_{12}$/Gd$_3$Ga$_5$O$_{12}$ interface to achieve the situation shown in Fig.~\ref{fig3_Ohnuma}.

\section{Discussion and conclusion \label{Sec:Conclusion}} 
The physics behind the predicted microwave emission are best understood in terms of the spin-pumping picture (see Fig.~\ref{fig1_Ohnuma}). First, recall that the spin accumulation can be regarded as dynamically induced paramagnons in the language of many-body theory~\cite{Hertz74}, and that the quantity $\sum_{\bf k} {\rm Im} \chi^{R}_{\bf k}(\omega)$ in Eqs.~(\ref{Eq:Dalpha01}) and (\ref{Eq:Ipump01}) denotes the density of states of the paramagnons~\cite{Doniach-text}. Then, without the temperature bias, a magnon carrying spin angular momentum $-\hbar$ pumps negative spin into $P$ in the form of paramagnon excitations. This always results in an enhancement of the Gilbert damping because the magnons lose the spin angular momenta [Fig.~\ref{fig1_Ohnuma}(a)]. On the other hand, when $P$ is heated, the pumped paramagnons tend to diffuse back into the cooler $F$, due to the drag effect of nonequilibrium phonons, which causes a forced spin backflow against the pumped spin current [Fig.~\ref{fig1_Ohnuma}(b)]. This spin backflow gives a negative contribution to the Gilbert damping constant in accordance with Eq.~(\ref{Eq:Dalpha02}) under the condition that $Q_{\rm ph}=-K_{\rm ph}(T_{\rm hot} - T_{\rm cold}) <0$. In this picture, the magnon instability in the limit of negligible intrinsic Gilbert damping constant $\alpha^{(0)}$ occurs when the sign of the spin-pumping current changes. 

Note that the present phenomenon differs from thermal spin-transfer torque in spin valve structures discussed in Refs.~\onlinecite{Hatami07} and \onlinecite{Slonczewski10}, as the spin valve structures are not required here. Furthermore, the phenomenon should be distinguished from a thermomagnonic analogue to Refs.~\onlinecite{Polianski04} and \onlinecite{Stiles04} in that the phonon-drag effect is the indispensable ingredient in the present theory. Moreover, the physics under discussion cannot be interpreted in terms of the spin-transfer torque exerted by a spin accumulation caused by the spin Seebeck effect, because such a process would yield a result proportional to $J_{\rm sd}^4$ ($J_{\rm sd}^2$ required for accumulating spin and the other $J_{\rm sd}^2$ for exerting torque) instead of $J_{\rm sd}^2$ in the present theory. Note also that, since we do not consider spin valve structures in this paper and there is only a uniform magnetization structure, we do not find any relevant processes in our calculation that are characterized by magnonic spin transfer torque. 

In conclusion, we have developed a theory for the thermal control of the Gilbert damping constant in a ferromagnet/paramagnet bilayer, and predicted that a microwave emission from the bilayer occurs at low temperature, where the participation of nonequilibrium phonons is remarkably enhanced. Since the structure of the bilayer is much simpler than that of spin valves, the present phenomenon extends the range of applications of spin caloritronics. 

\acknowledgments 
We would like to thank J. Ieda for useful comments and discussions. This work was financially supported by a Grant-in-Aid for Scientific Research on Innovative Area, ``Nano Spin Conversion Science'' (Grant No. 26103006) from MEXT, Japan, a Grant-in-Aid for JSPS Fellows from JSPS, Japan, and JSPS KAKENHI (Grant No. 15K05151).

\appendix

\section{Derivation of Eq.~(\ref{Eq:Dalpha02}) \label{Sec:Append1}} 
In this section, we derive Eq.~(\ref{Eq:Dalpha02}) in the main text. In the following, it is covenient to introduce a notation $d_i$ ($i=1,2,3$) to represent a temperature domain in Fig.~\ref{fig2_Ohnuma} having its local temperature $T_i$. 

We begin by deriving the expression of the phonon heat current $Q_{\rm ph}$ flowing into $P$. This quantity can be calculated in a manner similar to Ref.~\onlinecite{Maki65} using the definition $Q_{\rm ph}= \langle \frac{\partial}{\partial t} H^{(3)}_{\rm ph} \rangle$, where $\langle \cdots \rangle$ means the statistical average, and 
$H_{\rm ph}^{(3)} = \hbar \sum_{{\bf K}_3} \nu_{{\bf K}_3} b^\dag_{{\bf K}_3} b_{{\bf K}_3}$ 
is the Hamiltonian for phonons in domain $d_3$ with $b_{{\bf K}_3}$ being the phonon annihilation operator defined in the domain. To describe heat transfer across the domains, we need to introduce the interaction Hamiltonian 
\begin{eqnarray}
H^{(i,j)}_{\rm ph} &=& \hbar \sum_{{\bf K}_i, {\bf K}_j} 
\Omega^{(i,j)}({\bf K}_i-{\bf K}_j) B^{\dag}_{{\bf K}_i} B_{{\bf K}_j} + {\rm H.c.}, 
\end{eqnarray}
where $B_{{\bf K}_i}= b_{{\bf K}_i}+ b^\dag_{-{\bf K}_i}$, and $\Omega^{(i,j)}({\bf K})$ is the Fourier transform of $\Omega^{(i,j)} \sum_{{\bf r}_0 \in \text{interface}} v_0 \delta({\bf r} - {\bf r}_0)$ with $\hbar \Omega^{(i,j)}$ being the characteristic energy of the phonon transfer across the $d_i$/$d_j$ interface. 

Using the Heisenberg equation of motion for $H_{\rm ph}^{(3)}$ and then performing perturbation approach in terms of the interaction across the interface, we obtain 
\begin{eqnarray}
  Q_{\rm ph} &=& 
  \frac{\hbar L}{N_P N_F^2}\sum_{{\bf K}_1,{\bf K}_2,{\bf K}_3} \nu_{{\bf K}_3} \int_{\nu} 
  \, {\rm Im}D_{{\bf K}_1}(\nu) |D_{{\bf K}_2}(\nu)|^2 \nonumber \\ 
  &\times& 
     |{\rm Im}D_{{\bf K}_3}(\nu)| 
  \left[ \coth \big( \frac{\hbar \nu}{2 k_{\rm B}T_3} \big) 
    -\coth \big( \frac{\hbar \nu}{2 k_{\rm B}T_1} \big)\right], \nonumber \\
  \label{Eq:Qph00}
\end{eqnarray}
where $L=(\Omega^{(1,2)} \Omega^{(2,3)})^2 N_{\rm int}^{(1,2)} N_{\rm int}^{(2,3)}$ with $N_{\rm int}^{(i,j)}$ being the number of lattice sites at the $d_i$/$d_j$ interface, and we introduced the shorthand notation $\int_{\nu} = \int_{-\infty}^\infty \frac{d \nu}{2 \pi}$. We evaluate the integral over $\nu$ by picking up the dominant contribution that is proportional to the phonon lifetime $\tau_{\rm ph}$, and obtain 
\begin{eqnarray}
  Q_{\rm ph} &=& 
  \hbar L \tau_{\rm ph} \sum_{{\bf K}_1,{\bf K}_2,{\bf K}_3} 
  \nu_{{\bf K}_2} \, {\rm Im}D_{{\bf K}_1}(\nu_{{\bf K}_2}) 
     {\rm Im}D_{{\bf K}_3}(\nu_{{\bf K}_2}) \nonumber \\
     && \times \frac{(\frac{\hbar \nu_{{\bf K}_2}}{2 k_{\rm B}T})}
     {4 \sinh^2 (\frac{\hbar \nu_{{\bf K}_2}}{2 k_{\rm B}T})}
     \left( \frac{\Delta T}{T} \right), 
\end{eqnarray}
where the result was linearized with respect to $\Delta T$. 
After a little algebra, $Q_{\rm ph}$ is calculated to be 
\begin{eqnarray}
  Q_{\rm ph} &=& -K_{\rm ph} \Delta T, 
  \label{Eq:Qph01}\\
  K_{\rm ph} &=& \frac{\pi^3}{2 \hbar} L \tau_{\rm ph} 
  \frac{\rho^P_{\rm ph}(\nu_{P}) 
    \big( \rho^F_{\rm ph} (\nu_{P}) \big)^2 (k_{\rm B}T)^8}
       {(\hbar \nu_{P})^6 T} c_8, 
       \qquad   \label{Eq:Qph02}
\end{eqnarray}
where $\nu_{P}$ are the Debye frequency of $P$, 
$\rho^{P}_{\rm ph}(\nu)$ ($\rho^{F}_{\rm ph}(\nu)$) is the density of states of phonons in $P$ ($F$), and the constant $c_8$ is defined by 
\begin{eqnarray}
  c_8 &=& \int_0^{\hbar \nu_P/k_{\rm B}T} dx \frac{x^8}{4 \sinh^2(x/2)}. 
\end{eqnarray}

\section{Derivation of Eq.~(\ref{Eq:ISSE_micro}) \label{Sec:Append2}} 
In this section, we derive Eq.~(\ref{Eq:ISSE_micro}) in the main text. The spin current injected into $P$ via the spin Seebeck effect is given by~\cite{Adachi13} 
$I_{\rm S}^{\rm SSE}= \sum_{{\bf r}_i \in P} \langle \frac{\partial}{\partial t} s^z({\bf r}_i,t) \rangle$, 
where ${\bf s}({\bf r}_i,t)= N_P^{-1/2}\sum_{\bf k} {\bf s}_{\bf k}(t) e^{{\rm i}{\bf k}\cdot{\bf r}_i}$ 
is the spin accumulation in $P$. Using the Heisenberg Eq.~of motion for $s^z$ and assuming the steady state condition, this quantity can be calculated from 
$I_{\rm S}^{\rm SSE} = \sum_{{\bf k},{\bf q}} \int_\omega {\rm Re} \big[ J_{\rm sd}({\bf k}-{\bf q}) C^<_{{\bf k},{\bf q}}(\omega) \big]$, 
where $C^<_{{\bf k},{\bf q}}(\omega)$ is the Fourier transform of the interfacial Green's function $C^<(t,t')=-{\rm i}\langle a_{\bf q}(t') s_{\bf k}(t) \rangle$, and we introduced the shorthand notation $\int_\omega= \int_{-\infty}^{\infty} \frac{d \omega}{2 \pi}$. 

Now we focus on the process shown in Fig.~\ref{fig2_Ohnuma}(d). Following the procedure given in Ref.~\onlinecite{Adachi13b}, the corresponding spin current injection is represented as 
\begin{eqnarray}
  I_{\rm S}^{\rm SSE} &=& 
  \frac{J_{\rm sd}^2 S_0 N^{(2,3)}_{\rm int}}{\hbar^4 N_F N^2_P}
  \sum_{{\bf k},{\bf q},{\bf K}_3} \Upsilon^2_{{\bf K}_3} \int_\nu F_{{\bf k},{\bf q}}(\nu) 
      {\rm Im} \delta D^K_{{\bf K}_3} (\nu), \nonumber \\
      \label{Eq:ISSE01}
\end{eqnarray}
where $\delta D^K_{{\bf K}_3} (\nu)$ was defined in Eq.~(\ref{Eq:dD_K01}), and 
\begin{eqnarray}
  F_{{\bf k},{\bf q}} (\nu) &=& 
  \int_\omega |\chi^R_{\bf k}(\omega)|^2 {\rm Im}\chi^R_{{\bf k}_-}(\omega-\nu) 
      {\rm Im}G_{\bf q}(\omega) \nonumber \\
      &\times&
      \left[ \coth \big( \frac{\hbar (\omega-\nu)}{2 k_{\rm B} T} \big) 
         -\coth \big( \frac{\hbar \omega}{2 k_{\rm B}T} \big)\right]
\end{eqnarray}
with ${\bf k}_-$ being introduced below Eq.~(\ref{Eq:dSigma01}). We first use the relation $\coth(A-B)- \coth(A)= \sinh(B)/[\sinh(A)\sinh(A-B)]$ and then perform the integral over $\omega$ using ${\rm Im}G_{\bf q}(\omega)= - \pi \delta (\omega - \omega_{\bf q})$. Then, $F_{{\bf k},{\bf q}} (\nu)$ is calculated to be 
\begin{eqnarray}
  F_{{\bf k},{\bf q}} (\nu) &=& 
  - \frac{2 \pi k_{\rm B} T \nu \tau_{\rm sf}}{\hbar} 
  \frac{\big( \frac{\hbar (\omega_{\bf q}-\nu)}{2 k_{\rm B} T} \big)} 
         {\sinh \big( \frac{\hbar (\omega_{\bf q}-\nu)}{2 k_{\rm B} T}\big)}
         \frac{\sinh \big( \frac{\hbar \nu}{2 k_{\rm B} T}\big)}
              {\sinh \big( \frac{\hbar \omega_{\bf q}}{2 k_{\rm B} T}\big)} \nonumber \\
  &\times&  |\chi^R_{\bf k}(\omega)|^2 
  \frac{d}{d \omega} \left[ 
    \frac{{\rm Im} \chi^R_{\bf k}(\omega)}
    {\omega} \right]_{\omega=\omega_{\bf q}-\nu} , 
  \label{Eq:Fkq01}
\end{eqnarray}
where the derivative with respect to $\omega$ originates from the fact that only the even-in-$\nu$ component gives a non-vanishing contribution to $I_{\rm S}^{\rm SSE}$ because of the symmetry of the $\nu$ integral. 
Substituting Eq.~(\ref{Eq:Fkq01}) into Eq.~(\ref{Eq:ISSE01}) and performing the integral over $\nu$ by picking up the phonon poles, we obtain 
Eq.~(\ref{Eq:ISSE_micro}), where we have used an approximation 
$(\frac{\hbar (\omega_{\bf q}-\nu_{\bf K})}{2 k_{\rm B} T})/\sinh(\frac{\hbar (\omega_{\bf q}-\nu_{\bf K})}{2 k_{\rm B} T}) \approx 1$ because the dominant contribution comes from a region $\frac{\hbar (\omega_{\bf q}-\nu_{\bf K})}{2 k_{\rm B} T} \ll 1$.

\section{Comparison of the theory with experiments \label{Sec:Append3}}
Equation (\ref{Eq:Dalpha02}) means that the Gilbert damping constant is reduced when the paramagnet side is hotter, the sign of which is consistent with the experiment reported in Ref.~\onlinecite{Lu12}. For its order of magnitude estimation, we use Eq.~(\ref{Eq:Dalpha03}). Since we have $I_s^{\rm SSE}/\omega_0  \approx 2.0 \times 10^4$ at room temperature $T_{\rm R}$ under a temperature bias of $\Delta T=20$ K~\cite{Uchida12}, we have an estimate $\delta \alpha' \Delta T/[\alpha^{(0)}+ \delta \alpha] \approx 30$\%, where we used the fact that $\alpha^{(0)} \ll \delta \alpha $. This estimate is comparable to the experimentally observed reduction of $17$\%. 

On the other hand, one could attribute the observed reduction to a trivial effect of the temperature bias because there might be a temperature bias effect that is free from the phonon heat current $Q_{\rm ph}$ and comes solely from temperature dependence of material parameters in Eq.~(\ref{Eq:Dalpha01}). To see this, we first use the fact that the dominant temperature dependence comes from that of the paramagnetic spin susceptibility $\chi_{P}$~\cite{Kriessman54} and the size of the spin ($\propto$ saturation magnetization $M_{\rm s}$~\cite{Gilleo58}), where the former quantity is defined as a function of the temperature of $P$ (i.e., $T=T_P$) and the latter as a function of the temperature of $F$ (i.e., $T=T_F$). We next use the relation $T_F= T_{\rm R}+ \frac{1}{2}\Delta T$ and $T_P= T_{\rm R}- \frac{1}{2}\Delta T$ by setting $T_{\rm R}= \frac{1}{2}(T_F+ T_P)$, and calculate $\delta \alpha (T_{\rm R} +\Delta T)= \text{const}\times M_{\rm s}(T_{\rm R}+ \frac{1}{2}\Delta T) \times \chi_P(T_{\rm R}-\frac{1}{2}\Delta T)$. Then, we obtain $[\delta \alpha (T_{\rm R}+ \Delta T) -\delta \alpha (T_{\rm R})]/[\alpha^{(0)}+ \delta \alpha (T_{\rm R})]\approx \frac{1}{2}\frac{d}{dT} [\ln \chi_{P}(T) - \ln M_{\rm s}(T)]_{T=T_{\rm R}}\Delta T $, where we have again used the fact that $\alpha^{(0)} \ll \delta \alpha (T_{\rm R})$. This estimate gives at most a $2$\% reduction in $\alpha$ under a temperature bias of $\Delta T=20$ K, which is much smaller than the experimentally observed value. This means that Eq.~(\ref{Eq:Dalpha02}) is indeed the origin of the observed behavior.


\end{document}